# Electron-only Reconnection in Ion-scale Current Sheet at the Magnetopause

S. Y. Huang[1*], Q. Y. Xiong[1], L. F. Song[1], J. Nan[1], Z. G. Yuan[1], K. Jiang[1], X. H. Deng[2], and L. Yu[1]

[1]School of Electronic Information, Wuhan University, Wuhan, 430072, China

[2]Insititute of Space Science and Technology, Nanchang University, Nanchang, 330031, China

[*]Corresponding author: shiyonghuang@whu.edu.cn

**Abstract**

In the standard model of magnetic reconnection, both ions and electrons couple to the newly reconnected magnetic field lines and are ejected away from the reconnection diffusion region in the form of bi-directional burst ion/electron jets. Recent observations propose a new model: electron-only magnetic reconnection without ion coupling in electron-scale current sheet. Based on the data from Magnetospheric Multiscale (MMS) Mission, we observe a long extension inner electron diffusion region (EDR) at least 40 $d_i$ away from the X-line at the Earth's Magnetopause, implying that the extension of EDR is much longer than the prediction of the theory and simulations. This inner EDR is embedded in an ion-scale current sheet (the width of ~ 4 $d_i$, $d_i$ is ion inertial length). However, such ongoing magnetic reconnection was not accompanied with burst ion outflow, implying the presence of electron-only reconnection in ion-scale current sheet. Our observations present new challenge for understanding the model of standard magnetic reconnection and electron-only reconnection model in electron-scale current sheet.

## 1. Introduction

Magnetic reconnection is a widespread important physical process that allows the rapid energy conversion of magnetic field into the plasmas, resulting in the particle's acceleration/heating and the changing of magnetic field topology (Priest et al., 2000; Yamada et al., 2010). Magnetic reconnection is frequently observed or thought to play a major role in the astrophysical and space plasmas, such as solar flares, solar and stellar coronae, solar wind, planetary magnetosphere, the interplanetary space, the interstellar medium, neutron star, accretion disks, astrophysical jets, galaxy clusters and black holes (Priest et al., 2000; Øieroset et al., 2001; Deng et al., 2001; Lin et al., 2005; Vaivads et al., 2004; Huang et al., 2010; Yamada et al., 2010). The crucial region of magnetic reconnection can be divided into ion diffusion region

(IDR, where the ions are demagnetized) and electron diffusion region (EDR, where both the ions and the electrons are demagnetized) due to the different mass between the ion and the electron. Around or in the diffusion region, the significant phenomenon is Hall effect that includes Hall currents, bipolar Hall electric field pointing toward the center of the current sheet, and Hall quadrupolar out-of-plane magnetic field because of the relative motion between the ions and the electrons (Priest et al., 2000; Øieroset et al., 2001; Deng et al., 2001; Lin et al., 2005; Vaivads et al., 2004; Huang et al., 2010, 2012; Yamada et al., 2010). EDR, which is embedded in the ion diffusion region, can extend several $d_i$ along the outflow direction and develop two sub-structures, i.e., inner EDR and outer EDR. The inner EDR containing the X-line is the core region during the magnetic reconnection, which features intense electron currents, nonzero $\boldsymbol{E'} = \boldsymbol{E}+\boldsymbol{V_e}\times\boldsymbol{B}$, electron nongyrotropy or electron crescent distribution, super-Alfvénic electron flow, and electron dissipation $\boldsymbol{J}\bullet\boldsymbol{E'} > 0$, etc. (Zenitani et al., 2012; Hesse et al., 2013); while the outer EDR can extend tens of $d_i$ (where $d_i$ is ion inertial length) along the outflow direction, where the electrons are remained decoupled from the magnetic field and form a super-Alfvénic outflow jet, electron nongyrotropy, strong electron currents and $\boldsymbol{J}\bullet\boldsymbol{E'} < 0$ (Zenitani et al., 2012; Le et al., 2013).

The inner EDR has recently been *in-situ* identified at the terrestrial magnetopause and in the magnetotail by the unprecedented high-resolution measurements from the Magnetospheric Multiscale (MMS) mission (Burch et al., 2016; Zhou et al., 2017, 2019; Huang et al., 2018; Torbert et al., 2018; Fu et al., 2019). Outer EDR, with energy conversion from the particles to the fields ($\boldsymbol{J}\bullet\boldsymbol{E'}<0$), has been identified in the magnetosheath (Phan et al., 2007), in the magnetotail (Zhou et al., 2014) and at the magnetopause (Hwang et al., 2017). We should point out that such outer EDR is only an electron jet with the violation of electron frozen-in condition and has a negative $\boldsymbol{J}\bullet\boldsymbol{E'}$. Recently, an EDR with positive energy dissipation ($\boldsymbol{J}\bullet\boldsymbol{E'}>0$) extended 20 $d_i$ away from the X-line embedding in the burst ion outflow was reported in the downstream of magnetic reconnection at the magnetopause (Zhong et al., 2020).

Recent studies have presented reconnection with burst of electron jets but no ions accompanied at turbulent magnetosheath (Phan et al., 2018) and quasi-parallel shocks (Wang et al., 2019; Gingell et al., 2019), which challenges the standard model of EDR in the reconnection that the ions are ejected away from the diffusion region in the form of burst ion jets in the downstream. It is revealed that the electron-scale current sheet could also produce turbulent energy transformation and dissipation without ion participation during magnetic reconnection (Phan et al., 2018).

In this study, we report a textbook inner EDR emerging in an ion-scale current sheet (the thickness up to 4 $d_i$) at the magnetopause boundary layer, which is characterized by super-Alfvén electron jet, electron nongyrotropy, and positive energy dissipation, clear parallel electric field. This EDR with a thickness of ~0.53 $d_i$ is extended about 40 $d_i$ away from the X-line in the downstream of magnetic reconnection but without burst ion outflow. Our observations demonstrate a new feature of reconnection in space, i.e., electron-only reconnection in the ion-scale current sheet, which is different from the traditional magnetic reconnection model, and challenges the previous observations as well.

## 2. Event overview

The overview observations from 13:31:10 to 13:34:00 UT on September 07, 2015 when the MMS were located at ~ [3.85, 10.9, -0.12] $R_E$ ($R_E$ is the Earth's radius) are shown in Figure 1. The magnetic field measured by the fluxgate magnetometer (FGM) instrument (Russell et al., 2016), the electric field from the electric double probes (EDP) instruments (Lindqvist et al., 2016), and the particle data measured by the fast plasma investigation (FPI) instrument (Pollock et al., 2016) onboard MMS in burst mode are used in this study. MMS traveled through firstly magnetosphere (positive $B_z$ in Figure 1(a), low speed plasma flow, high temperature and low density in Figure 1(b)-1(e)), then crossed magnetopause boundary layer, and finally entered the magnetosheath (negative $B_z$ in Figure 1(a), high speed flow, low temperature and high density in Figure 1(b)-1(f)). These two black dashed lines mark the magnetopause boundary layer which is characterized by the mixed particles: high energy particles from the magnetosphere and low energy particles from the magnetosheath (Figure 1(g) and 1(h)). One reconnecting current sheet marked by yellow shadow has most intense current density (Figure 1(i)), up to ~ 2 $\mu A/m^2$, which will be investigated in the following part.

Figure 2 displays the detailed observations of the current sheet from 13:32:26 UT to 13:32:28 UT. All vectors are presented at LMN coordinate system derived by the minimum variance analysis (MVA) (Sonnerup et al., 1998), where $L$ direction is along with the magnetic field line trend, $N$ points to the inflow direction and $M$ component completes the orthogonal coordinate system. Three uniformed vectors of LMN coordinates in the GSE coordinate, are $L$ = [-0.206, -0.216, 0.955], $M$ = [0.089, -0.975, -0.202], and $N$ = [0.975, 0.043, 0.220], respectively. This crossing of the current sheet embedded in steady ion flow (Figure 2(b)) is characterized by the reversal of $B_L$ from positive to negative (Figure 2(a)), accompanied by a bipolar signature of $B_M$ component relative to the guide field (~ -8 nT) (Figure 2(a)), tripolar signature of current $J_L$ component (Figure 2(d)), large fluctuations in electron velocity (Figure 2(c)) and intense electric field (Figure 2(e)). Combing the tripolar variation in $J_L$, and bipolar variation in $B_M$, one can infer that MMS detected one reconnection diffusion region with well-known Hall current, Hall quadrupolar out-of-plane magnetic field (small guide field ~ -8 nT here). Due to the large convective ion flow $V_i$, the convection term will dominate the electric field. Thus, the convective term $V_i \times B$ removed from the electric field $E_N$, i.e., $(E+V_i \times B)_N$ is shown in Figure 2(f). One can see that $(E+V_i \times B)_N$ has a bipolar variation from positive to negative except one small pulse during the crossing of the current sheet, indicating that $(E+V_i \times B)_N$ points toward the center of the current sheet. We also calculate the different terms of the general Ohm's law. It can be seen that Hall term $(J \times B)_N$ can well balance the electric field $(E+V_i \times B)_N$, which means that the bipolar change of $(E+V_i \times B)_N$ is Hall electric field caused by the Hall term. The electron velocity $V_{eL}$ and $V_{eM}$ are up to -300 km/s and 400 km/s respectively after subtracting the background flow, which are much larger than the local Alfvén speed $V_A$ ~ 126 km/s, implying that MMS detected a super-Alfvénic electron flow in this diffusion region. In addition, MMS also measured one peak in electron density and the increase in electron temperature dominated by parallel temperature. It is interesting that the non-zero electric field $\boldsymbol{E'}$ in the electron frame (Figure 2(g)), large parallel electric field (up to 5 mV/m in Figure 2(h)), and very strong energy dissipation from the fields to the plasmas ($\boldsymbol{J}\bullet\boldsymbol{E'}>0$, up to 7 nW/m$^3$ in Figure 2(k)) are observed during this crossing. All these features suggest the existence of inner EDR in this reconnection diffusion region. It is noticeable that the ion bulk velocity at LMN coordinate does not have obvious increase signature as Figure 1b-1c shows. According to the electron motion during the time interval, the background flow is against to the reconnection outflow direction (center vertical dashed line), which should cause similar decrease of total velocity for the electrons and ions. Thus, the slight increase of ion bulk velocity in Figure 2b and Figure 1c should be the increase of background flow, not the signature of ion outflow. Therefore, only the electron outflow is detected during the current sheet crossing, implying that this event could be categorized as electron-only reconnection.

Figure 3 shows the observations of the electron anisotropy and energy dissipation from four MMS spacecraft. It can be seen that all four MMS spacecraft captured the EDR with peaks in positive energy dissipation ($\boldsymbol{J}\bullet\boldsymbol{E'}>0$ in Figure 3(b)), agyrotropy $\sqrt{Q}$ and $A\phi_e$ (Figure 3(c)-3(d)), the gain energy per cyclotron period $\varepsilon_e$ (Figure 3(e)), and relative strength of electric and magnetic force in the bulk electron rest frame $\delta_e$ (Figure 3(f)) which are marked by four vertical dashed lines, implying that four MMS spacecraft successively crossed the EDR. One should point that these peaks are not in the center of the current sheet (i.e., $B_L=0$), but located at the positive $B_L$ region, indicating that the inner EDR has a deflection due to the presence of a guide field, which has been predicted in the previous simulations (Le et al., 2013) and observed by the Cluster (Zhou et al., 2014). This influence can also explain that the center of current sheet and the center of EDR are inconsistent which showed in Figure 2a.

We perform the Timing analysis on the magnetic field and obtain the moving speed $V_n \sim 234$ km/s along the direction $n$ = [0.0288 0.1493 -0.9884] in LMN coordinates. The thickness of the current sheet is estimated as $V_n \times dt \sim$ 234 km/s × 0.7 s = 164 km, or 161 km along $N$ direction (where $dt$ is the average duration of the current sheet crossing for four MMS spacecraft), about 4 $d_i$ or 161 $d_e$ ($d_i \sim$ 41 km and $d_e \sim$ 1 km are the ion and electron inertial lengths respectively based on background ion density $n_i \sim$ 30 cm$^{-3}$). This indicates that MMS encountered a thick current sheet or ion-scale current sheet. In addition, we notice that four MMS spacecraft successively crossed the current sheet, firstly MMS1, then MMS2 and MMS3, and finally MMS4, thus we also can use the positions of the MMS to estimate the thickness of the current sheet. When the MMS1 (MMS2) was in the center of the current sheet, MMS2 (MMS4) was at the edge of the current sheet; when MMS3 (MMS4) was in the center of the current sheet, MMS1 (MMS2) was at the edge of the current sheet (seen $B_L$ in Figure 3(a)). The separations of MMS1-MMS2, MMS1-MMS3, MMS2-MMS4 along $N$ direction are 67 km, 98 km, and 83 km, respectively. Thus, the thickness can be estimated as 2×(67+ 98+ 83 +83)/4 ~ 165.5 km in $N$ direction, which is very close to the estimated thickness derived by the Timing analysis. Moreover, based on the average duration (0.095 s) of positive $\mathbf{J} \cdot \mathbf{E}'$, the estimated thickness of the inner EDR along $N$ direction is about 22 km, i.e., ~ 0.54 $d_i$ or 22 $d_e$.

The trajectory of the MMS crossing the reconnection region is illustrated in Figure 4. In order to determine where the EDR is, we identify the separatrix and the center of EDR using $B_M$-$B_g$ = 0 (where guide field $B_g \sim$ -8 nT) marked by three vertical dashed lines in Figure 2. The EDR extension slightly deviates from the center of current sheet due to the presence of guide field (Le et al., 2013). The distance between two separatrices and the center of EDR is estimated as 33 km and 55 km respectively based on the Timing analysis, contributing to ~2 $d_i$ width between two separatrices. Given the hypothesis that the separatrices are straight lines start from X line as well as the magnetic field lines are parallel to the separatrices nearby, one can obtain the cone angle of the spacecraft crossing point at separatrices from the following equation: $\tan\theta \sim \frac{\delta}{D} \sim \frac{B_N}{B_L}$, where $\delta$ is the distance between the separatrix and the center of EDR, $D$ is the extension of EDR from the X-line. The cone angle $\theta$ can be derived by the magnetic field in $L$ and $N$ direction at the intersection of the trajectory of MMS and the separatrices. Thus, the cone angles $\theta$ are ~1.19° and ~1.94° corresponding to the lower and upper crossing point at separatrices respectively (shown in Figure 4). Based on the triangle theory, the extension length of EDR is estimated to 1604 km and 1621 km away from the X-line resulting from the distance as 33 km and 55 km between the center of EDR and lower and upper separatrices respectively. In roughly, therefore, the EDR extension from the X-line in the downstream is at least ~ 40 $d_i$, which is the first time *in-situ* observation of inner EDR for such a long extension in space. In addition, the average reconnection rate $R$ = 0.021 ~ 0.034 calculated by the equation given in Liu et al. (2017) and Nakamura et al. (2018), consistent with the previous predictions and observations (Xiao et al., 2007; Liu et al., 2017; Chen et al., 2019; Zhong et al., 2020).

## 3. Conclusion

Thanking for the unprecedented high-resolution data from the MMS mission, the inner EDR is successfully and definitely identified at the magnetopause (Burch et al., 2016; Zhou et al., 2017), in the magnetotail (Huang et al., 2018; Torbert et al., 2018; Chen et al., 2019; Zhou et al., 2019) by electron nongyrotropy or electron crescent distribution, strong energy dissipation $\boldsymbol{J}\bullet\boldsymbol{E'}$ > 0, super-Alfvénic electron flow, parallel electric field, and electron demagnetization, etc. In addition, the outer EDR with electron demagnetization and super-Alfvénic outflow jet is also identified in the magnetosphere (Xiao et al., 2007; Chen et al., 2019). In present study, we identify a reconnection diffusion region with **well-defined** Hall electromagnetic field and Hall current at the magnetopause boundary layer. Further analysis shows that the EDR embedded in this diffusion region has a long extension, at least ~ 40 $d_i$ away from X-line in the downstream within the ion-scale current sheet (4 $d_i$). This EDR does not belong to outer EDR, but consistent with the signatures of inner EDR. Recently, Phan et al. (2018) have shown an electron-only reconnection in electron-scale current sheets. However, it is surprising that there is not burst ion outflow (Figure 2b) even in such a thick ion-scale current sheet in our case, implying that this inner EDR occurs during electron-only magnetic reconnection in an ion-scale current sheet. This implies that energy transformation and dissipation without ion participation during magnetic reconnection could also occur in an ion-scale current sheet, like in the electron-scale current sheet.

Our observations reveal a new feature of the EDR in magnetic reconnection, which challenges the understanding of standard EDR in magnetic reconnection and recent electron-only reconnection in electron-scale current sheet, and gives new pictures for the magnetic reconnection. The event presented in this study could be the textbook for identifying the details of an electron-only type magnetic reconnection. Recently the possible mechanisms of electron-only reconnection formation have been proposed. It is suggested that the electron-only reconnection is the early phase of an ion scale reconnection (Wang et al., 2020). Meanwhile, the strong external driver (Lu at al., 2020) could also be the reasons forming the electron-only reconnection. It is still, however, doubtful about the applicability of these mechanisms on the ion-scale current sheet circumstances. It would be testified through simulations in future. Our results could also shed new lights on fundamental understanding of the reconnection process in the astrophysical and space plasmas.

**Acknowledgement**

This work was supported by the National Natural Science Foundation of China (41874191, 42074196, 41925018), and the National Youth Talent Support Program. SYH appreciates the fruitful discussions with Dr. M Zhou. The datasets analyzed in present study are publicly available from the MMS Science Data Center (https://spdf.gsfc.nasa.gov/pub/data/mms).

**References**

Burch, J. L., Torbert, R. B., Phan, T. D., et al. 2016b, *Sci*, 352, aaf2939

Chen, L.- J., Wang, S., Hesse, M., et al. 2019, *GeoRL*, 46, 6230–6238

Deng, X., & Matsumoto, H. 2001, *Natur*, 410, 557–560

Fu, H. S., Cao, J. B., Cao, D., et al. 2019, *GeoRL*, 46, 48–54

Gingell, I., Schwartz, S. J., Eastwood, J. P., et al. 2019, *GeoRL*, 46, 1177-1184

Hesse, M., Aunai, N., Zenitani, S., et al. 2013, *PhPl*, 20, 061210

Huang, S. Y., Zhou, M., Sahraoui, F., et al. 2010, *JGRA*, 115, A12211

Huang, S. Y., Vaivads, A., Khotyaintsev, Y. V., et al. 2012, *GeoRL*, 39, L11103

Huang, S. Y., Jiang, K., Yuan, Z. G., et al. 2018, *ApJ*, 862, 144

Hwang, K.- J., Sibeck, D. G., Choi, E., et al. 2017, *GeoRL*, 44, 2049– 2059

Le, A., Egedal, J., & Ohia, O. 2013, *PhRvL.*, 110, 135004

Lin, J., Ko, Y. K., Sui, L., et al. 2005, *ApJ*, 622, 1251

Lindqvist, P.-A., Olsson, G., Torbert R. B., et al. 2016, *SSRv*, 199, 137–165

Liu, Y. H., Hesse, M., Guo, F., et al. 2017, *PhRvL*, 118, 085101

Lu, S., Wang, R., Lu, Q., et al. 2020, *NatComm*, 11, 5049

Nakamura, T. K. M., Genestreti, K. J., Liu, Y.-H., et al. 2018, *JGRA*, 123, 9150– 9168

Øieroset, M., Phan, T. D., Fujimoto, M., et al. 2001, *Natur*, 412, 414–417

Phan, T. D., Drake, J. F., Shay, M. A., et al. 2007, *PhRvL*, 99, 225,002

Phan, T. D., Eastwood, J. P., Shay, M. A., et al. 2018, *Natur*, 557, 202–206

Pollock, C., Moore, T., Jacques, A., et al. 2016, *SSRv.*, 199, 331–406

Priest, E., & T. Forbes (2000), Magnetic Reconnection: MHD Theory and Applications, Cambridge Univ. Press, Cambridge, U. K.

Russell, C. T., Anderson, B. J., Baumjohann, W., et al. 2014, *SSRv.*, 199, 189–256

Sonnerup, B. U. O., & Scheible, M. 1998, in Analysis Methods for Multi-Spacecraft Data, ed. G. Paschmann & P. Daly (Noordwijk: ESA), 185

Torbert, R. B., Burch, J. L., Phan, T. D., et al. 2018, *Sci*, 362, 1391

Vaivads, A., Khotyaintsev, Y., André, M., et al. 2004, *PhRvL*, 93, 105001

Wang, R. S., Lu, Q. M., Lu, S., et al. 2020, *GeoRL*, 47, e2020GL088761

Wang, S., Chen, L.- J., Bessho, N., et al. 2019, *GeoRL*, 46, 562-570

Xiao, C., Pu, Z. Y., Wang, X. G., et al. 2007, *GeoRL*, 34, L01101

Yamada, M., & Ji, H. 2010, *Proceedings of the International Astronomical Union*, 6(S274), 10-17.

Zenitani, S., Hesse, M., Klimas, A., et al. 2012, *PhRvL*, 106, 195003

Zhong, Z. H., Zhou, M., Tang, R. X., et al. 2020, *ApJL*, 892, 1

Zhou, M., Deng, Tang, X., R., et al. 2014, *JGRA*, 119, 1541-1548

Zhou, M., Berchem, J., Walker, R. J. J., et al. 2017, *PhRvL.*, 119, 055101

Zhou M., Deng, X. H., Zhong, Z. H., et al. 2019, *ApJ*, 870, 1

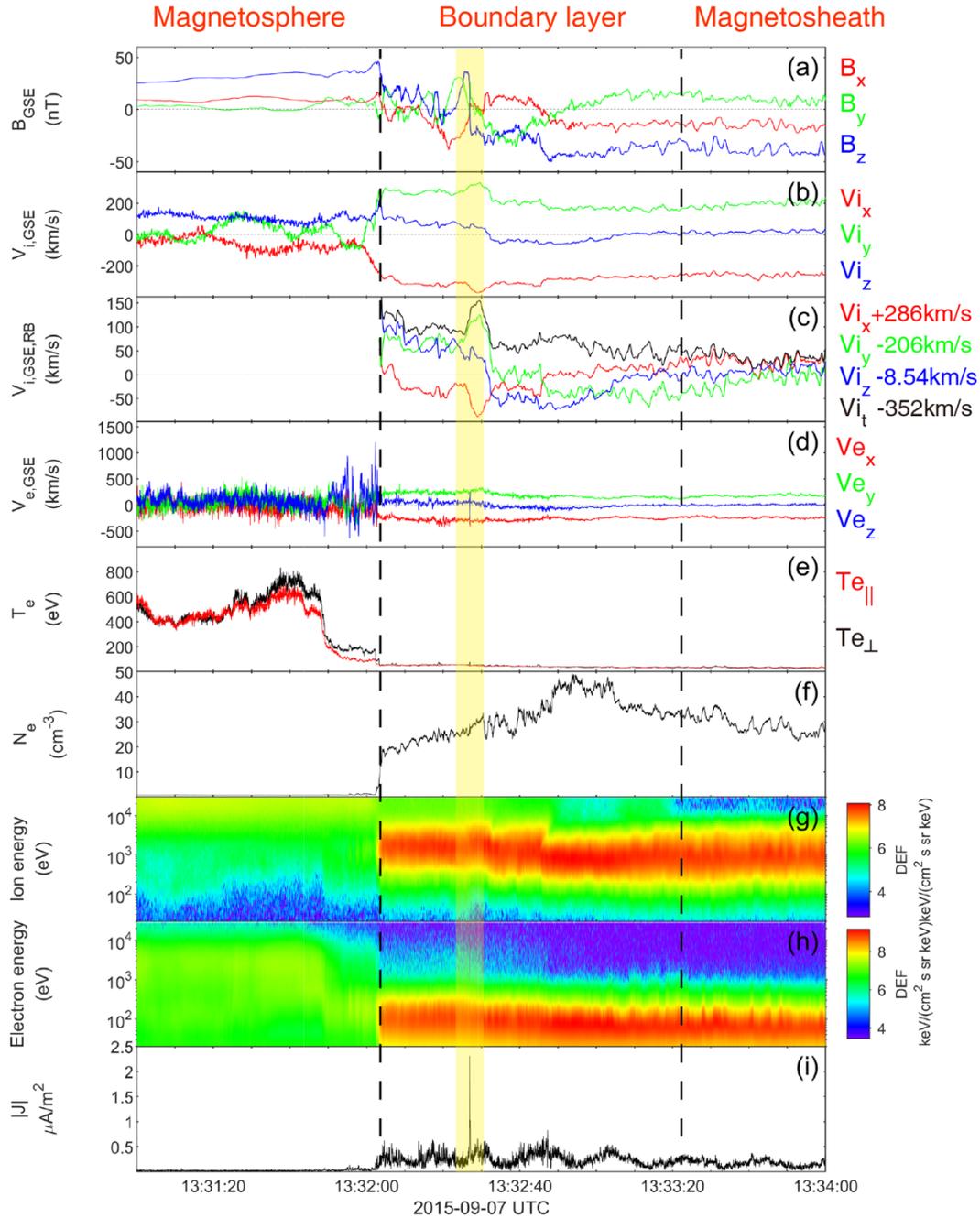

**Figure 1**. Overview observations on September 07, 2015. (a-d) Three components of magnetic field, ion bulk velocity, ion bulk velocity which has been removed background flow and electron velocity in the GSE coordinates, respectively. (d) Electron temperature. (e) Electron density. (f-g) Ion and electron energy fluxes. (h) the total current density. Two black dashed lines mark the magnetopause boundary layer. The yellow shallow highlights the interval of the reconnecting current sheet.

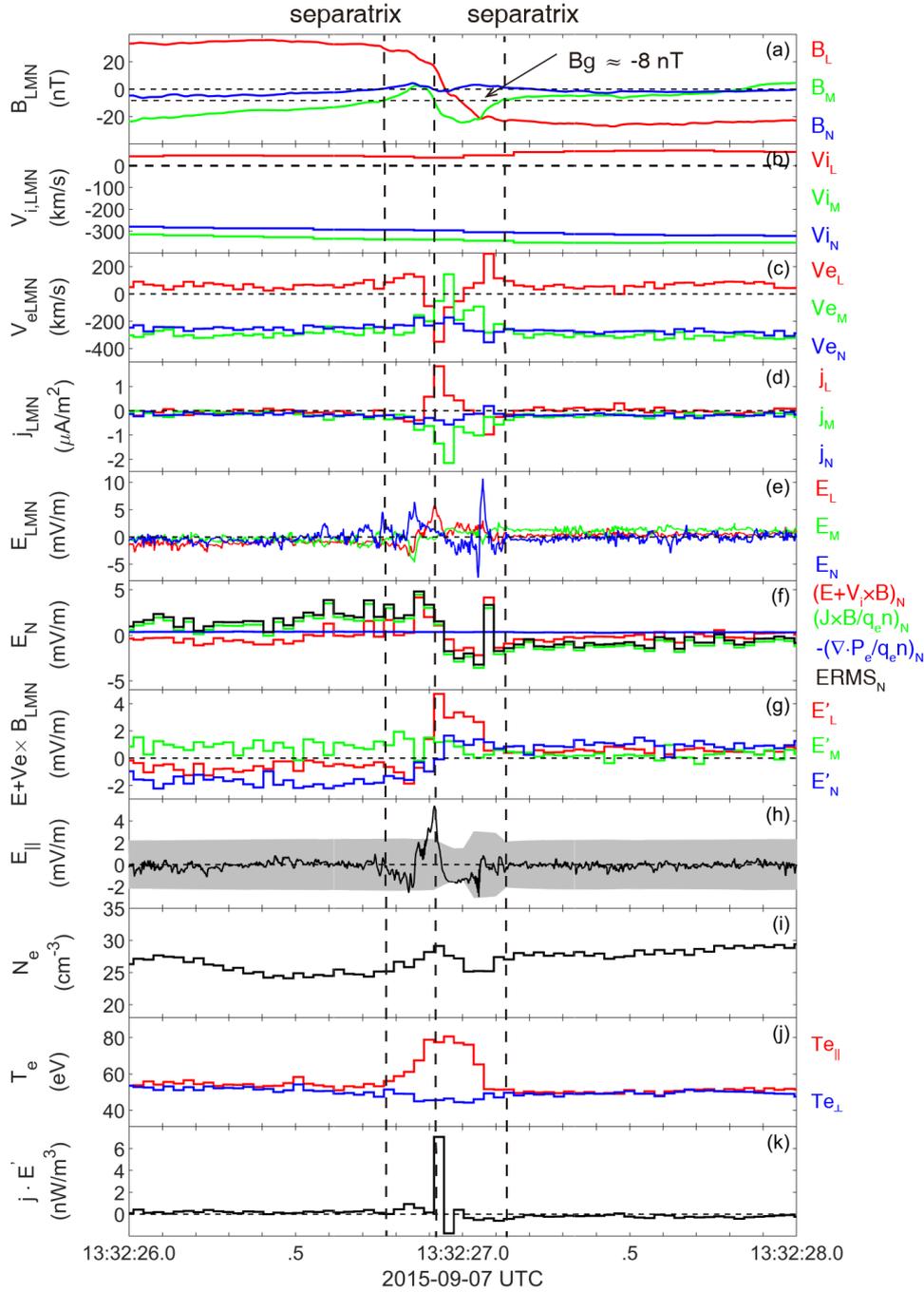

**Figure 2.** Detailed observations of the thick current sheet. (a) Magnetic field, (b) ion velocity, (c) electron velocity, (d) current density, (e) electric field, (f) electric field in the ion frame ($E + V_i \times B$)$_N$, Hall term and electron pressure gradient term, (g) electric field in the electron frame $E + V_e \times B$, (h) parallel electric field, (i) electron density, (j) electron parallel (red) and perpendicular (blue) temperature, (k) $J \cdot E'$ (where $E' = E + V_e \times B$, and $J$ is calculated with plasma moments). All vectors are showed in LMN coordinate system. Three vertical black dashed lines mark the lower separatrix, the center of EDR, and upper separatrix.

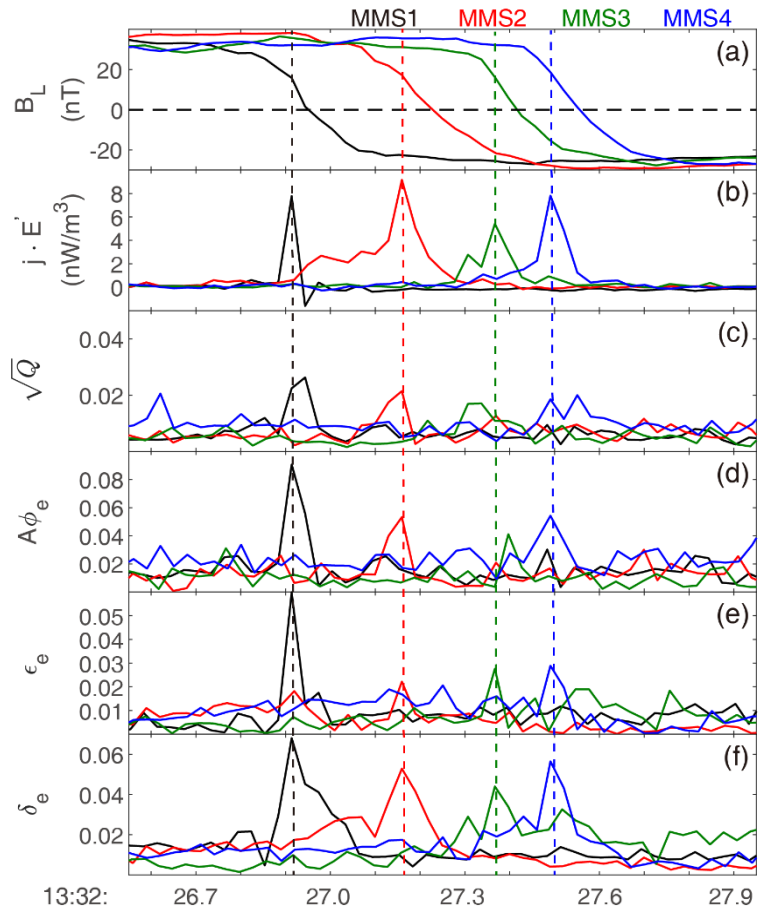

**Figure 3**. EDR successively observed by four MMS spacecraft. (a) $L$ component of magnetic field. (b) Energy dissipation $\boldsymbol{J} \cdot \boldsymbol{E'}$. (c-d) Electron agyrotropy using two methods: $\sqrt{Q}$ considers the full agyrotropy with all components of electron pressure tensor while $A\phi_e$ only measures the components perpendicular to magnetic field. (e) Energy gain per cyclotron period of electron $\varepsilon_e$. (f) Relative strength of electric and magnetic force in the bulk electron rest frame $\delta_e$.

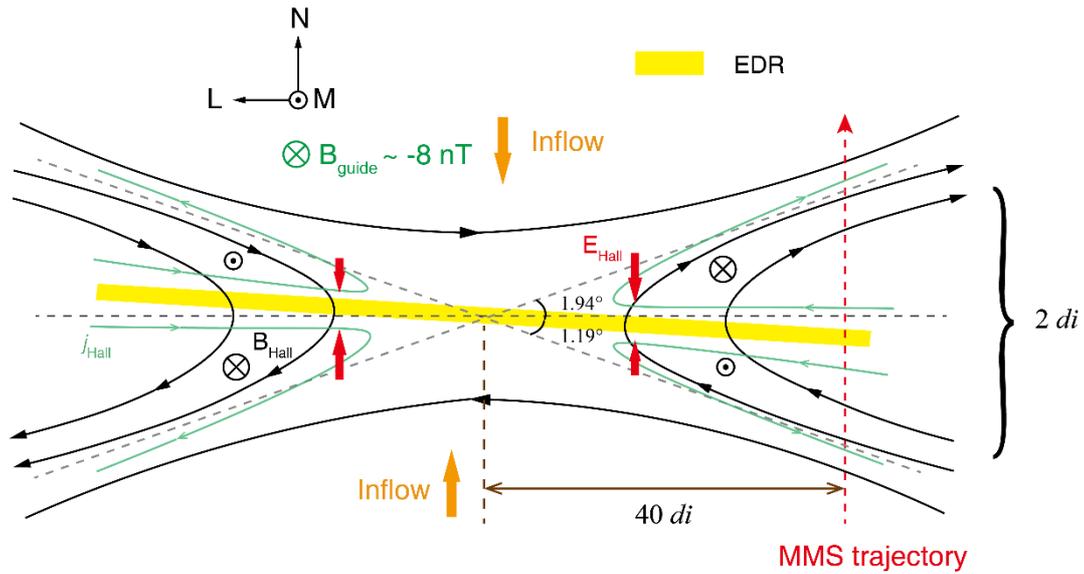

**Figure 4**. Illustration of reconnection site in *L-N* plane. MMS (whose trajectory is denoted by magenta dashed line) cross the EDR (yellow region) along the N direction. Green curves represent the Hall current sheet system. Red arrows point to the Hall electric field. Orange arrows show the inflow of ions and electrons. Two dashed gray lines define the separatrices. The cone angles between the separatrix and EDR extension are 1.94° (upper) and 1.19° (lower). The horizontal dashed line is the center of current sheet (i.e., the region with $B_L=0$). The width between upper separatrix and lower separatrix along the trajectory of MMS is ~2 $d_i$.